\documentclass{JHEP3}
\usepackage{epsfig}

\newcommand{\be}{\begin{equation}}
\newcommand{\ee}{\end{equation}}

\title{Computation of the string tension in three dimensional 
Yang-Mills theory using large $N$ reduction}
\author{ Joe Kiskis
\\ Department of Physics, University of California,
Davis, CA 95616, USA\\E-mail: \email
{jekiskis@ucdavis.edu} }
\author{Rajamani Narayanan
\\Department of Physics, Florida International University, Miami,
FL 33199, USA\\E-mail: \email{rajamani.narayanan@fiu.edu}}

\abstract {
We numerically compute the string tension in the large $N$
limit of three dimensional Yang-Mills theory using Wilson
loops. Space-time loops are formed as products of smeared
space-like links and unsmeared time-like links. We use 
continuum reduction and both
unfolded and folded Wilson loops in the analysis.
}


\preprint{}

\begin{document}

\section{Introduction}

The method of large $N$ continuum 
reduction~\cite{Narayanan:2003fc,Kiskis:2003rd}
 for $SU(N)$ gauge
theory allows for the calculation of the infinite volume, infinite $N$
limit of certain physical quantities using volumes reduced to a small
physical size. 
Numerical
estimates~\cite{Narayanan:2003fc,Kiskis:2003rd} of the physical 
critical size above which continuum reduction
holds 
indicate that this method can be used to produce practical
results.
The chiral condensate~\cite{Narayanan:2004cp}
and the pion decay constant\cite{Narayanan:2005gh}
were calculated in the
large $N$ limit in four dimensions using continuum reduction. 
In this paper, we show that the
method can be extended beyond bulk quantities and that it also
produces reliable results for quantities with space-time dependence
such as the heavy quark potential, from which the string tension can
be extracted.

A precise calculation of the string tension in three dimensional
$SU(N)$ gauge theories has been performed 
with $N$ up to 8 on large lattices~\cite{Bringoltz:2006zg}.
In this paper, we present a complementary calculation with $N=47$ on
$5^3$ lattices using continuum reduction. 
The calculation of Ref.~\cite{Bringoltz:2006zg} used
correlation functions of smeared Polyakov loops to extract the string 
tension. After extrapolating to $N=\infty$ and to
the continuum, the result was
\be
\frac{\sqrt\sigma}{g^2N} = 0.1975 \pm 0.0002 - 0.0005
\label{bt1}
\ee
where $g$ is the gauge coupling.
This has to be compared with the analytical calculation 
in~\cite{Karabali:1998yq}, namely, $\frac{1}{\sqrt{8\pi}}\approx0.1995$.
Although the two results are not in perfect agreement, 
the main observation
is that the approximations used in the analytical calculation
are very well motivated. 

The string tension for $SU(N)$ as per the 
analytical calculation~\cite{Karabali:1998yq}
is 
\be
\sigma = \frac{ g^2 c_A}{2\pi} \frac{ g^2 c_F}{2} =
\left[g^2N\right]^2 \frac{1}{8\pi} \left[ 1-\frac{1}{N^2}\right]  .
\ee
$c_A$ and $c_F$ are the quadratic Casimirs in the adjoint and
fundamental representation, respectively.
The mass parameter $\frac{ g^2 c_A}{2\pi}$ naturally enters
the analytical calculation, and the second factor
$\frac{ g^2 c_F}{2}$ arises from the Wilson loop
operator in the fundamental representation.

The numerical computation in~\cite{Bringoltz:2006zg}
shows that the agreement gets better as one gets closer
to $N=\infty$. Like the analytical result, the numerical
result also shows a correction from $N=\infty$ that goes
like $\frac{1}{N^2}$, but the coefficient of $\frac{1}{N^2}$
is not the same for the numerical and the analytical computations.

Since the numerical and the analytical results are close to
each other even for $N=2$ (less than $4\%$), we can use
the analytical formula to get a feel for the finite $N$ corrections
to the infinite $N$ result.
Therefore, we expect the finite $N$ corrections to be smaller than
the error in~Eqn(\ref{bt1}) if $N > 32$.

In this paper, we use continuum 
reduction~\cite{Narayanan:2003fc,Kiskis:2003rd}
to directly compute the $N=\infty$ limit of the string tension
by working at large enough $N$ so that the finite $N$ corrections
are smaller than the numerical errors. We find that
\be
\frac{\sqrt\sigma}{g^2N} = 0.1964 \pm 0.0009
\label{thisresult}
\ee
This result and that of (\ref{bt1}) 
are consistent at the level of their
one sigma errors. This level of agreement is, in turn, consistent with
neither the large $N$ extrapolation of Ref.~\cite{Bringoltz:2006zg} 
nor the volume
reduction of the present calculation having unexpected errors. While
both of the numerical results lie below the analytical estimate, the
discrepancy is relatively small. Thus the numerical evidence that the
analytical result is an excellent first approximation that captures
much of the physics remains strong.

The paper is organized as follows. 
We explain how we use
smeared Wilson loops to compute the string tension in 
Section~\ref{details}. The lattice results for the string tension
along with the continuum extrapolation are also presented
in this section. An intermediate step in our calculation is
the dimensionless ground state string energy $m(k)$. 
In Section~\ref{string}, we show results for
$m(k)$ at one fixed lattice coupling to illustrate its behavior
as a function of $k$ and how it is used to extract the string
tension. We also show that $m(k)$ is unaffected by the smearing
parameter. We illustrate the extraction of $m(k)$ at one fixed
coupling in Section~\ref{mass}. Here we show how the smearing
parameter affects the overlap with the ground state.
The main result in this paper is obtained using $N=47$.
We show that the finite $N$ and finite volume 
corrections are small at this
value of $N$ in Section~\ref{finiteN}. We explain why
this method is preferred over the Creutz ratio 
in Section~\ref{crratio}.

\section{String tension using Wilson loops and continuum
reduction}\label{details}

Consider $SU(N)$ Yang-Mills theory on a periodic lattice
with the standard Wilson gauge action. 
The method of~\cite{Bringoltz:2006zg} is to
measure the string tension using correlations of Polyakov loops
with separation $t$ that wind around a space direction.
Continuum reduction~\cite{Narayanan:2003fc,Kiskis:2003rd}
implies that the
large $N$ Yang-Mills theory in a continuum box of size
$l^3$ is independent of $l$ as long as $ l > l_c = 1/T_c$
with $T_c$ being the deconfining temperature.
One should be able to compute expectation values
of Wilson loops of arbitrary
size on an $l^3$ continuum box using folded Wilson loops and
extract the string tension. To implement this approach to the 
three-dimensional Yang-Mills theory string tension, we use the 
following procedure:
\begin{itemize}
\item We fix the lattice size to $L^3$. We use $L=5$ for
the most part and only use $L=4$ to verify reduction.
\item We fix $N$ so that finite $N$ corrections are small.
We set $N=47$ and show using one instance that
finite $N$ corrections are small at $N=47$.
\item We pick an appropriate range of lattice coupling
$b=\frac{1}{g^2N}$.
\begin{itemize}
\item $b$ cannot be too small since we have to be away
from the bulk transition on the lattice associated
with the development of gap in the eigenvalue distribution
of the plaquette operator~\cite{Bursa:2005tk}. Therefore,
we pick $b\ge 0.6$.
\item $b$ cannot be too big since we have to be below
the deconfining transition for $L=5$. Therefore, we pick
$b \le 0.8$~\cite{Narayanan:2007ug}.
\end{itemize}
\item We use smeared space-like links and unsmeared
time-like links.
\item We use the tadpole improved coupling $b_I=be(b)$
to set the scale
and consider $K \times T$ Wilson loops $W(K,T)$ with 
$1.5 < \frac{K}{b_I},\frac{T}{b_I} < 12.5$.
This amounts to expectation values of Wilson loops that range from
$0.82$ to $2\cdot 10^{-4}$.
\item Keeping $K$ fixed, we fit 
\be
ln W(k,t) = - a - m(k) t;
\ee
where
$ k=\frac{K}{b_I}$ and $t=\frac{T}{b_I}$ are the dimensionless
extent in the space and time direction respectively.
$m(k)$ is the dimensionless ground state energy.
This fit assumes that there is a perfect overlap with the
ground state. Note that 
$a$ should be zero since $W(k,0)=1$. Any small deviation from
zero seen in the fit is due to the contribution from
excited states. This can be seen by noting that 
$W(k,t)$ should behave as $e^{-a} e^{-m(k) t} + (1-e^{-a}) e^{-m_1(k)t}$
with $m_1(k) > m(k)$. Then, 
$ln W(k,t) = -a - m(k) t + ln \left[ 1 + \left(e^a-1\right)
e^{-[m_1(k) - m(k)] t}\right]$. The last term is numerically 
insignificant in the range of $t$ being considered. 
\item Finally, $m(k)$ is fit to
 $\sigma b_I^2 k + c_0b_I + \frac{c_1}{k}$.
The combination $\sqrt{\sigma}b_I$ is plotted as a function
of $b_I^{-2}$. We expect lattice spacing
effects to lead off as $b_I^{-2}$ in Yang-Mills theories and
this is indeed the case in Fig.~\ref{stringcont}. The
continuum limit extracted from this figure was quoted 
in~Eqn.(\ref{thisresult}).
\end{itemize}

The use of smeared links improves the
measurement of Wilson loops. 
They enhance the overlap of the space-like sides of the Wilson loops 
with the ground state. This increases the signal relative to the 
fluctuations and simplifies the $t$ 
behavior of the loops~\cite{Teper:1998te}. One step in the iteration
takes one from a set $U^{(i)}_k (x_1,x_2,t)$ to a set 
$U^{(i+1)}_k (x_1,x_2,t)$,
by the following equation:
\begin{eqnarray}
X^{(i+1)}_1 (x_1,x_2,t)&=&
(1-f) U^{(i)}_1 (x)
+\frac{f}{2}
U^{(i)}_2 (x_1,x_2,t)U^{(i)}_1 (x_1,x_2+1,t)
\left[U^{(i)}_2 (x_1+1,x_2,t)\right]^\dagger\cr
&&+\frac{f}{2}
\left[U^{(i)}_2 (x_1,x_2-1,t)\right]^\dagger U^{(i)}_1 (x_1,x_2-1,t)
U^{(i)}_2 (x_1+1,x_2-1,t)\cr
X^{(i+1)}_2 (x_1,x_2,t)&=&
(1-f) U^{(i)}_2 (x)\cr
&&+\frac{f}{2}
U^{(i)}_1 (x_1,x_2,t)U^{(i)}_2 (x_1+1,x_2,t)
\left[U^{(i)}_1 (x_1,x_2+1,t)\right]^\dagger\cr
&&+\frac{f}{2}
\left[U^{(i)}_1 (x_1-1,x_2,t)\right]^\dagger
U^{(i)}_2 (x_1-1,x_2,t)
U^{(i)}_1 (x_1-1,x_2+1,t)\cr
U^{(i+1)}_k (x_1,x_2,t)&=&X^{(i+1)}_k (x_1,x_2,t)
\frac{1}{\sqrt{[X^{(i+1)}_k (x_1,x_2,t)]^\dagger 
X^{(i+1)}_k (x_1,x_2,t)}};\ k=1,2
\label{smear}
\end{eqnarray}
Note that time-like links, $U_3(x_1,x_2,t)$, are not smeared.
Also note that smearing only involves space-like staples.
There are two parameters, namely, the smearing factor $f$
and the number of smearing steps $n$. Only the product
$\tau=fn$ matters, and $f$ plays the role of a discrete smearing
step. For a given $\tau$, the overlap of the
smeared loop with the ground state does not depend on $f$
as long as it is small. But the overlap of the
smeared loop with the ground state does depend upon $\tau$.
We set the value of the smearing parameter
to $\tau=2.5$ by choosing $f=0.1$ and $n=25$.
To study the effect of varying $\tau$, we also consider
$\tau=1.25$ ($f=0.05$ and $n=25$) at one coupling.

{
 \EPSFIGURE
{stringcont.eps, width=\textwidth}
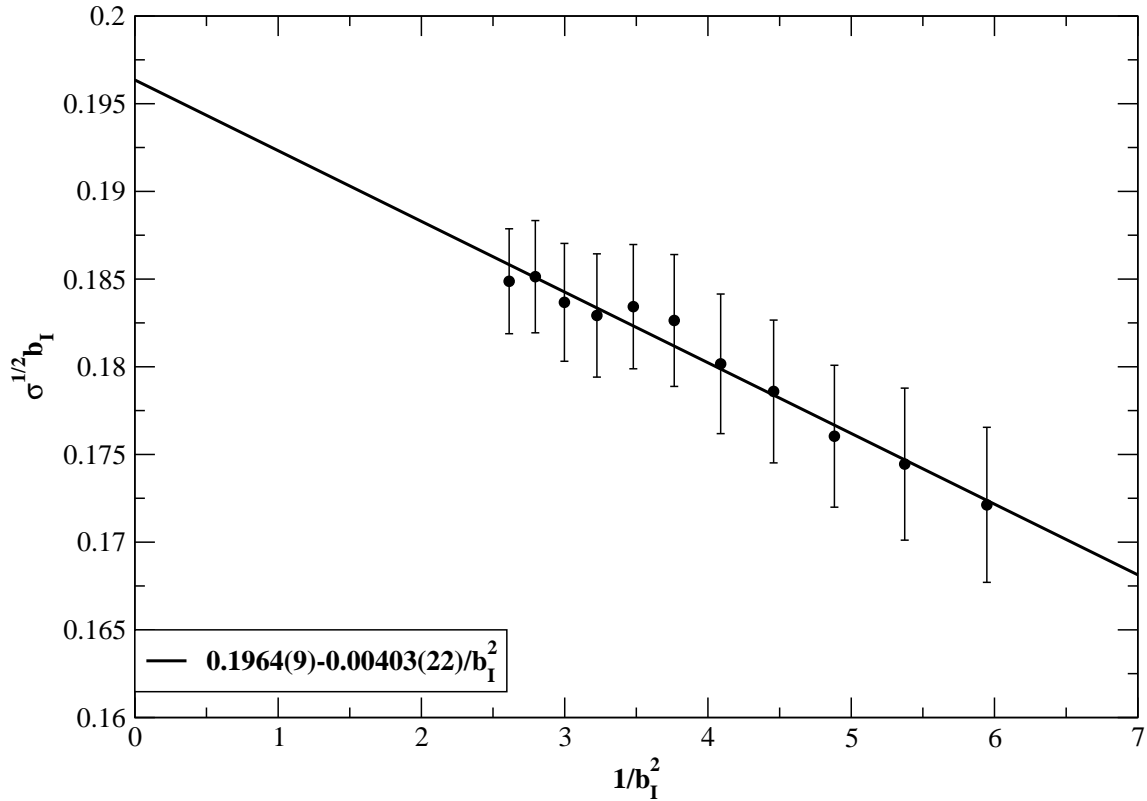
{ The string tension is plotted as a function of the
lattice spacing $b_I^{-1}$. The fit is an extrapolation to
the continuum.\label{stringcont}}
}

\section{Extraction of string tension}\label{string}

$SU(N)$ gauge fields were generated on a $5^3$ periodic lattice
using the standard Wilson action. One gauge 
field update of the whole lattice~\cite{Kiskis:2003rd} is
one Cabibbo-Marinari heat-bath update of the whole lattice
followed
by one $SU(N)$ over-relaxation update of the whole lattice.
A total of $1500$ such updates were used to achieve
thermalization. Measurements were separated by $10$ such
updates and all estimates are from a total of $832$ such
measurements. Errors in all quantities at a fixed $b$
and $N$ were obtained by jackknife with
single elimination.

{
\EPSFIGURE
{mass.eps, width=\textwidth}
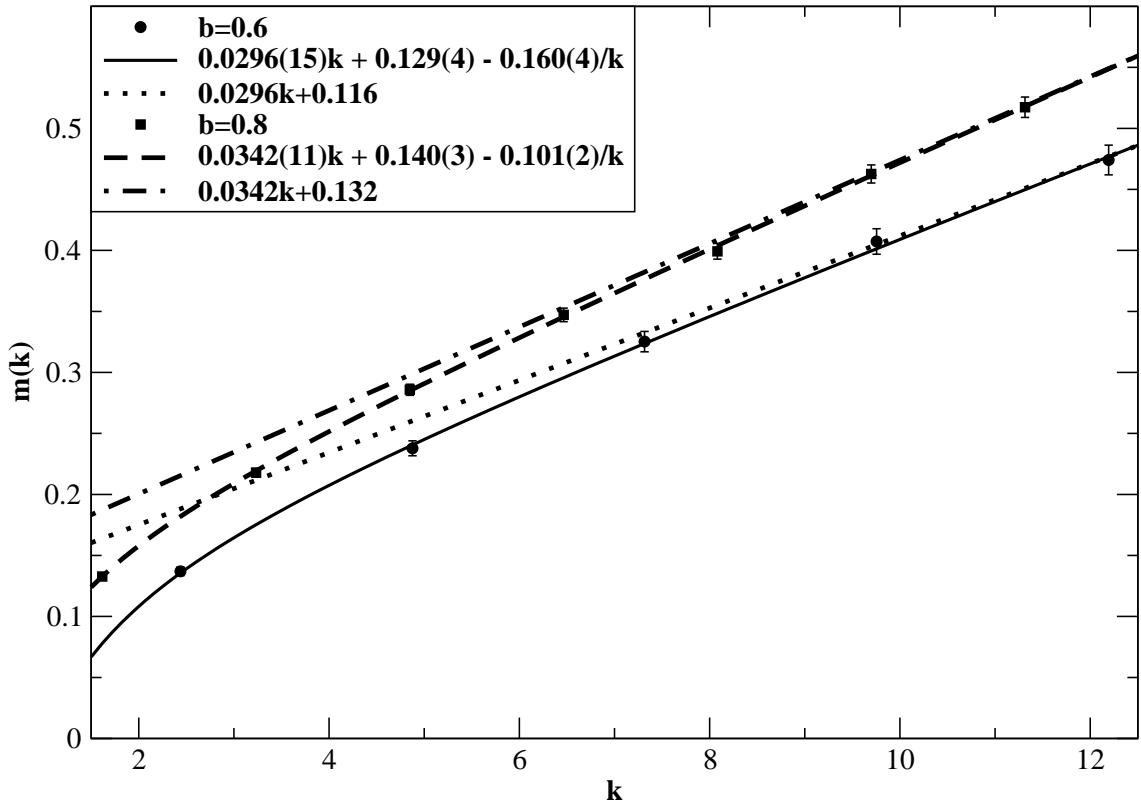
{ The ground state energy $m(k)$ as a 
function of $k$ for the coarse and fine lattice
spacings considered here.
\label{tension}
}}

The ground state energy $m(k)$ obtained as a function
of $k=\frac{K}{b_I}$ is fit to
\be
m(k) = \sigma b_I^2 k + c_0 b_I + \frac{c_1}{k}\label{stringfit}
\ee
We expect $\sigma b_I^2$ to approach a finite value in the 
continuum limit ($b_I\to\infty$). 
The same is expected for $c_1$. For large, unsmeared or 
symmetrically smeared Wilson loops with $t \gg k$, the 
universal value is 
$-\frac{\pi}{24} \approx -0.13$~\cite{Luscher:1980fr} 
rather than the value $-\frac{\pi}{6}$ 
for Polyakov loops~\cite{de Forcrand:1984cz} that was seen
in~\cite{Bringoltz:2006zg}.
The $c_0 b_I$ term is present
due to the perimeter divergent contribution, and therefore
it is expected to logarithmically diverge in the continuum
limit. Since, we do not smear the time-like gauge fields,
the divergence in this term is not tamed. 

The method will encounter difficulties in
extracting the physically relevant string tension
from Eq.(\ref{stringfit}) if $c_0b_I$ is large.
However, because we do not go to very weak couplings, we see in
Fig.~\ref{tension} that $c_0b_I$ is not too large.
It is not
necessary to go to weaker couplings since the string
tension computed at the couplings we have chosen
can be used to get a good estimate of the string tension
in the continuum limit as is evident in Fig.~\ref{stringcont}.

The three parameter fit of $m(k)$ as a function of $k$
is shown in Fig.~\ref{tension}. The fit has two degrees
of freedom at the coarse lattice spacing of $b=0.6$ and
has four degrees of freedom at the fine lattice spacing
of $b=0.8$.
 As mentioned before, errors in $\sigma b_I^2$,
$c_1$, and $c_0b_I$ are obtained by jackknife with
single elimination. Unlike the estimate of the leading
coefficient $\sigma b_I^2 $, the estimates of the sub-leading ones
are not as reliable. Consider the {\sl dotted} line
and {\sl dot-dashed} line in Figure~\ref{tension}.
The constant term in each of these lines is obtained
by evaluating $c_0b_I + c_1/k$ at $k=12.5$. The
coefficient of the linear term in each is set to the same value
as the one in the full three parameter fit. A comparison
of the {\sl dotted} line with the {\sl solid} line
and a comparison of the {\sl dot-dashed} line with the
{\sl dashed} line shows that the $1/k$ term becomes
relevant only if $k < 7$. There are only two data points
at $b=0.6$ and three data points at $b=0.8$ where the
$1/k$ effect is significant. As such, we do not expect
our estimate of $c_1$ and $c_0b_I$ to be as reliable
as the estimate of $\sigma b_I^2$.

The behaviors of $c_0b_I$ 
and $c_1$ as a function
of $b_I^{-2}$ are shown in Fig.~\ref{linear}
and Fig.~\ref{luescher}. Between these two terms,
$c_0b_I$ is the dominant one. 
The rise in $c_0b_I$ at smaller $b_I^{-2}$ is consistent with
the presence
of a $\ln b_I$ term from a perimeter
divergence. 
The estimate of $c_1$ is least reliable
since it is dominated by the other two terms in the
fit of $m(k)$ as a function of $k$. A fit of
$c_1$ as shown in Fig.~\ref{luescher} is consistent
with the existence of a continuum limit. 
One should note that the number of degrees of freedom
in the three parameter fit of $m(k)$ increases
as $b_I$ decreases and this will have an effect in
the determination of the sub-leading terms. We believe
this is reason the fit of $c_1$ versus $b_I$ does not
pass through all the data points. 
Since we do space-like but not time-like smearing and since our loops 
do not generally have $t \gg k$, it not surprising to see a 
result that disagrees with the universal value but has the same sign and 
order of magnitude.

\section{Extraction of $m(k)$}\label{mass}

{
\vskip 0cm \DOUBLEFIGURE
{linear.eps, width=0.45\textwidth}
{luescher.eps, width=0.45\textwidth}
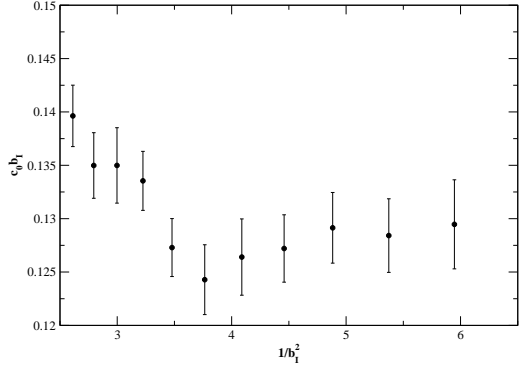
{The behavior of the coefficient $c_0b_I$ in the fit of
$m(k)$ vs $k$ is consistent with the presence of a 
$\ln b_I$ term due to the perimeter divergence.
\label{linear}
}
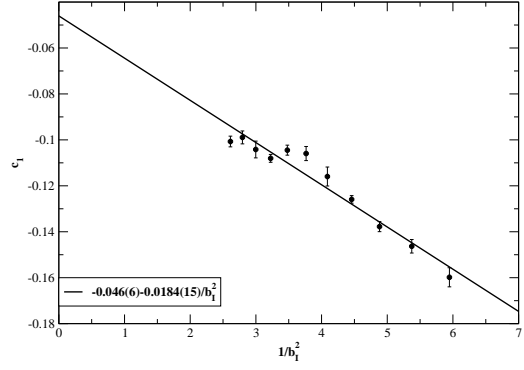
{The behavior of the coefficient $c_1$ in the fit of
$m(k)$ vs $k$ shows the existence of a continuum limit.
\label{luescher}
}
}
{
 \DOUBLEFIGURE
{wloop0.1.eps, width=0.45\textwidth}
{wloop0.05.eps, width=0.45\textwidth}
{ 
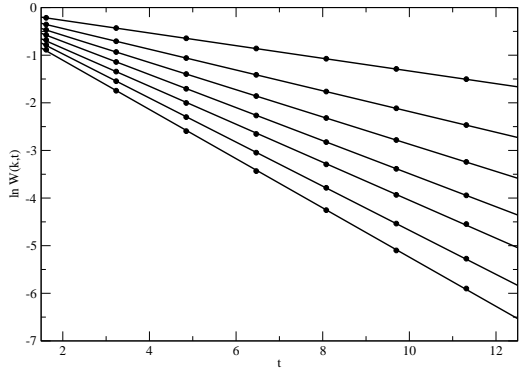
Plot of $\ln W(k,t)$ as a function of $t$ for seven
different values of $k$ at $b=0.8$ with $\tau=2.5$. \label{wloop1}}
{ 
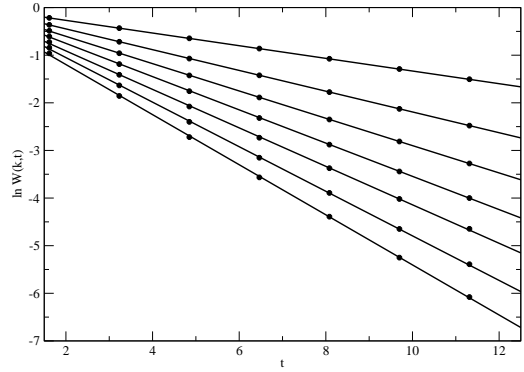
Plot of $\ln W(k,t)$ as a function of $t$ for seven
different values of $k$ at $b=0.8$ with $\tau=1.25$.\label{wloop2}}
}

The dimensionless ground state energy $m(k)$ is extracted
at a fixed $k$ by fitting $\ln W(k,t)$ to $-a-m(k)t$ as
discussed in Sec.~\ref{details}.
While $m(k)$ should be independent of the smearing parameter 
$\tau=fn$,
the value of $a$ 
is expected to depend $\tau$. 

We will use $b=0.8$ as the coupling to illustrate the extraction
of $m(k)$.
Figure~\ref{wloop1} and Fig.~\ref{wloop2} show the performance
of the fit for two different values of $\tau$, namely, $2.5$
and $1.25$ respectively. The {\sl solid circles} show the data
points without errors. The {\sl solid lines} show the fit of
the data. Seven values of $t$ were used to fit the data at one
$k$, and data at seven different values of $k$ were fitted.
This amounted to all Wilson loops from $1\times 1$ to $7\times 7$
on the $5^3$ lattice. The set of thermalized configurations
used at $\tau=2.5$ is statistically independent from the
set used at $\tau=1.25$. The fit parameters are shown in
Table~\ref{tab1} and Table~\ref{tab2}. Only the average
values of the fit parameters are listed. 

Investigation of Table~\ref{tab1} and Table~\ref{tab2}
shows that $m(k)$ does not depend on $\tau$. There is
a small difference in the two values of $m(k)$ at a fixed
$k$ for the two different values of $\tau$ if $k$ is
large. But Fig.~\ref{massf} shows that this difference
is within errors. Furthermore, the fitted values of
$\sigma b_I^2$ for the two different values of $\tau$
are the same within errors.

{
\TABULAR
{cccccccc}
{$k$ & 1.62 & 3.23 & 4.85 & 6.47 & 8.08 & 9.70 & 11.31 \cr
$a$ & 0.001 & 0.003 & 0.009 & 0.019 & 0.055 & 0.047 & 0.071 \cr
$m(k)$ & 0.133 & 0.218 & 0.286 & 0.347 & 0.399 & 0.464 & 0.517 \cr
}
{Fit parameters corresponding to the fit $\ln W(k,t) = -a -m(k) t$
for seven different values of $k$ at $b=0.8$ with $\tau=2.5$.
\label{tab1}}}

The values of $a$ in Table~\ref{tab1} and Table~\ref{tab2}
do show a variation with $\tau$ and $k$. Since a smaller
value of $\tau$ implies less smearing, the overlap with
the ground state is less for smaller $\tau$, and this
results in a larger value of $a$ at smaller $\tau$.
The value of $a$ is very close to zero for small $k$
indicating excellent overlap with the ground state for
the chosen value of $\tau$. As $k$ increases, the length
of the loop increases and the perimeter divergence has
a stronger effect. This results in a larger value of
$a$ as $k$ increases at a fixed $\tau$.
{
\TABULAR
{cccccccc}
{$k$ & 1.62 & 3.23 & 4.85 & 6.47 & 8.08 & 9.70 & 11.31 \cr
$a$ & 0.002 & 0.012 & 0.029 & 0.054 & 0.102 & 0.114 & 0.144 \cr
$m(k)$ & 0.133 & 0.218 & 0.287 & 0.349 & 0.404 & 0.468 & 0.526 \cr
}
{Fit parameters corresponding to the fit $\ln W(k,t) = -a -m(k) t$
for seven different values of $k$ at $b=0.8$ with $\tau=1.25$.
\label{tab2}}}

{\vskip 2cm
 \EPSFIGURE
{massf.eps, width=\textwidth}
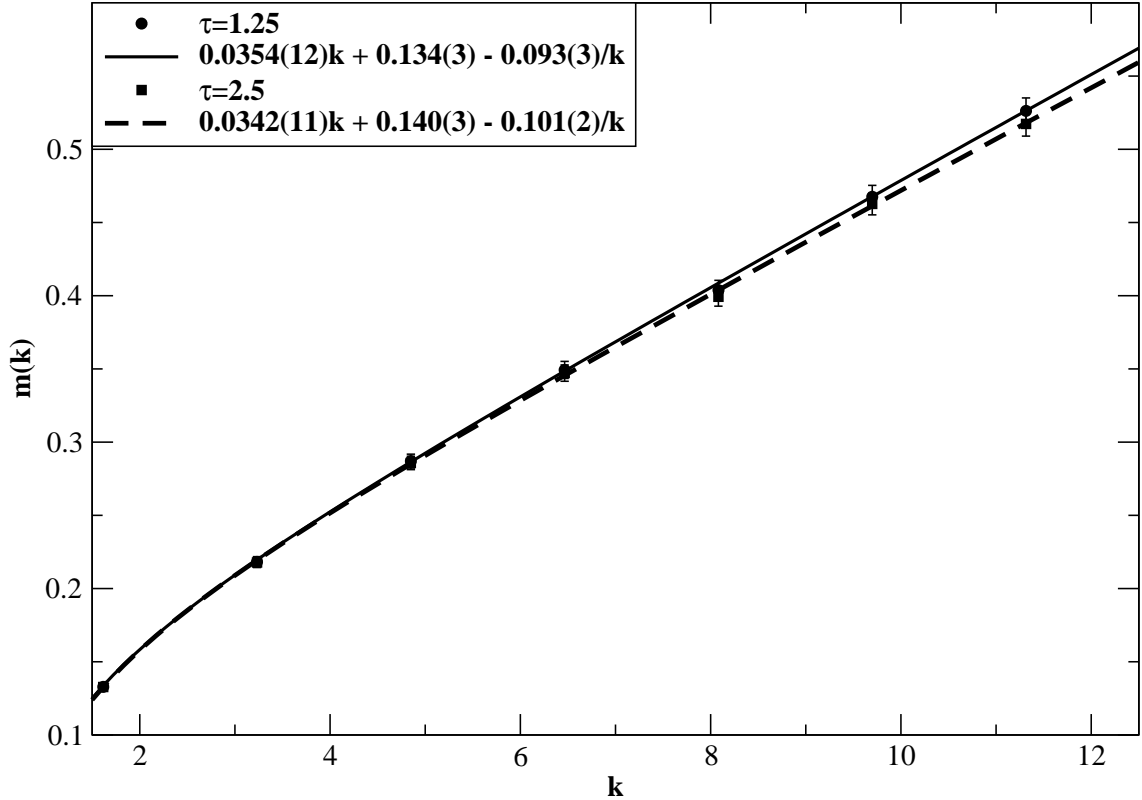
{ The ground state energy $m(k)$ as a 
function of $k$ for two different values of the smearing
parameter at $b=0.8$.
\label{massf}
}}

\section{Finite $N$ effects}\label{finiteN}

Two issues need to be addressed with the analysis performed
so far. We have fixed our value of $N$ assuming finite $N$
effects are small. If $N$ is not large enough, finite $N$
effects need to be addressed. In addition, we also have
to address finite volume effects since continuum reduction
is valid only in the $N\to\infty$ limit.

We expect $m(k)$ to have a fixed limit as $N\to\infty$
at a fixed $k$, $L$, $b$ and $\tau$. Indeed, this is
the case as shown in Fig.~\ref{massn} where the results
for $m(k)$ as a function of $k$ are shown for $b=0.8$
with $\tau=2.5$ on $5^3$ lattice. All three fit parameters
are consistent within errors all the way from $N=23$ to
$N=47$. The only glitch one sees is at $k\approx 8$.
This corresponds to $K=kb_I=5$, which is
the linear extent of the lattice. One can argue that there are
larger finite $N$ effects at strong coupling
for $K=L$.
Since the fit of $m(k)$ involves several values of $k$,
the larger effect at this particular value of $k$ is
diminished in the extraction of $\sigma b_I^2$.

{\vskip 2cm
 \EPSFIGURE
{massn.eps, width=\textwidth}
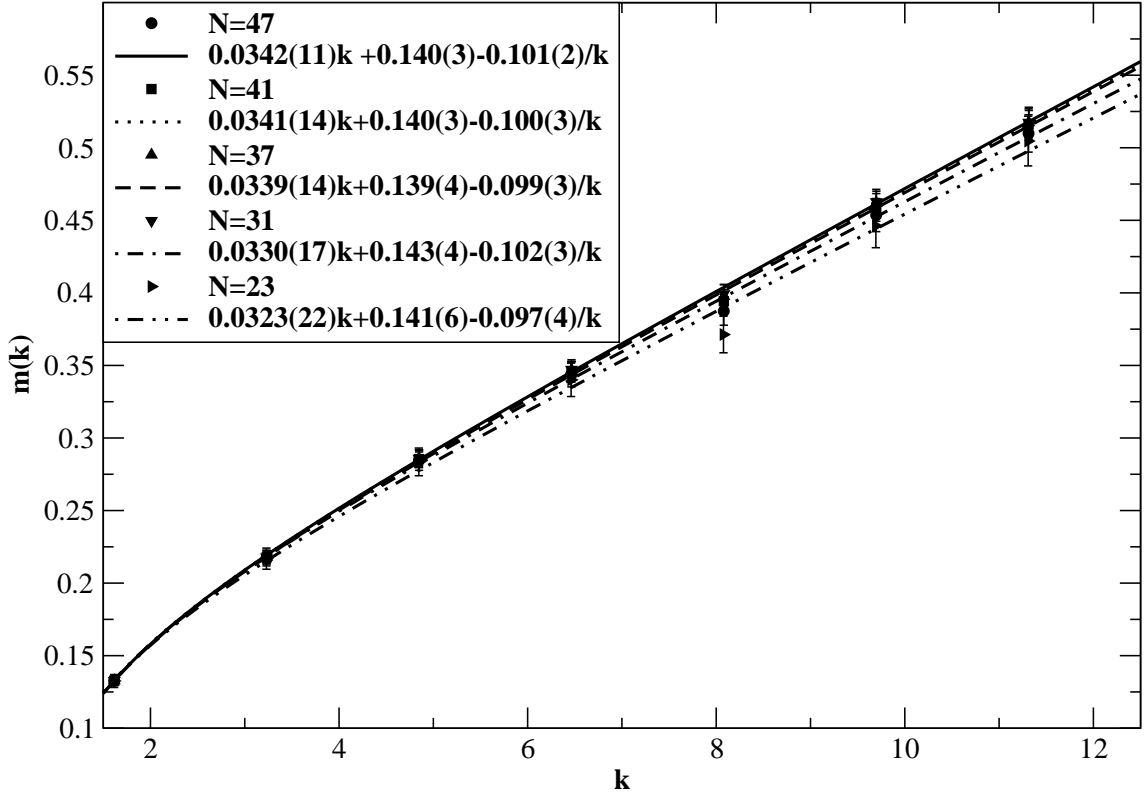
{ The ground state energy $m(k)$ as a 
function of $k$ for five different values of $N$
at $b=0.8$.
\label{massn}
}}

Since finite $N$ effects can be ignored at $N=47$, we also
expect there to be no appreciable finite volume effects at
this value of $N$. This point is illustrated in Fig.~\ref{massl}
where the result for $m(k)$ is plotted at $b=0.6$ and $\tau=2.5$
on $4^3$ and $5^3$ lattice. We used $b=0.6$ for this comparison
since we have to be in the confined phase on $4^3$ lattice.
Figure~\ref{massl} shows that the two values of $m(k)$ 
at a fixed $k$ are consistent with each other within errors.
The same is the case for the fit parameter $\sigma b_I^2$.
This is not the case for $c_1$ and $c_0 b_I$, and this is
probably due to a three parameter fit using only five data points.
Sub-leading coefficients are expected to depend sensitively on
the data points. Since we are primarily concerned with the
value of the string tension in this paper and since all our
results are based on data taken on $5^3$, we expect the
final result to be free of finite $N$ and finite $L$ errors.

{\vskip 2cm
\EPSFIGURE
{massl.eps, width=\textwidth}
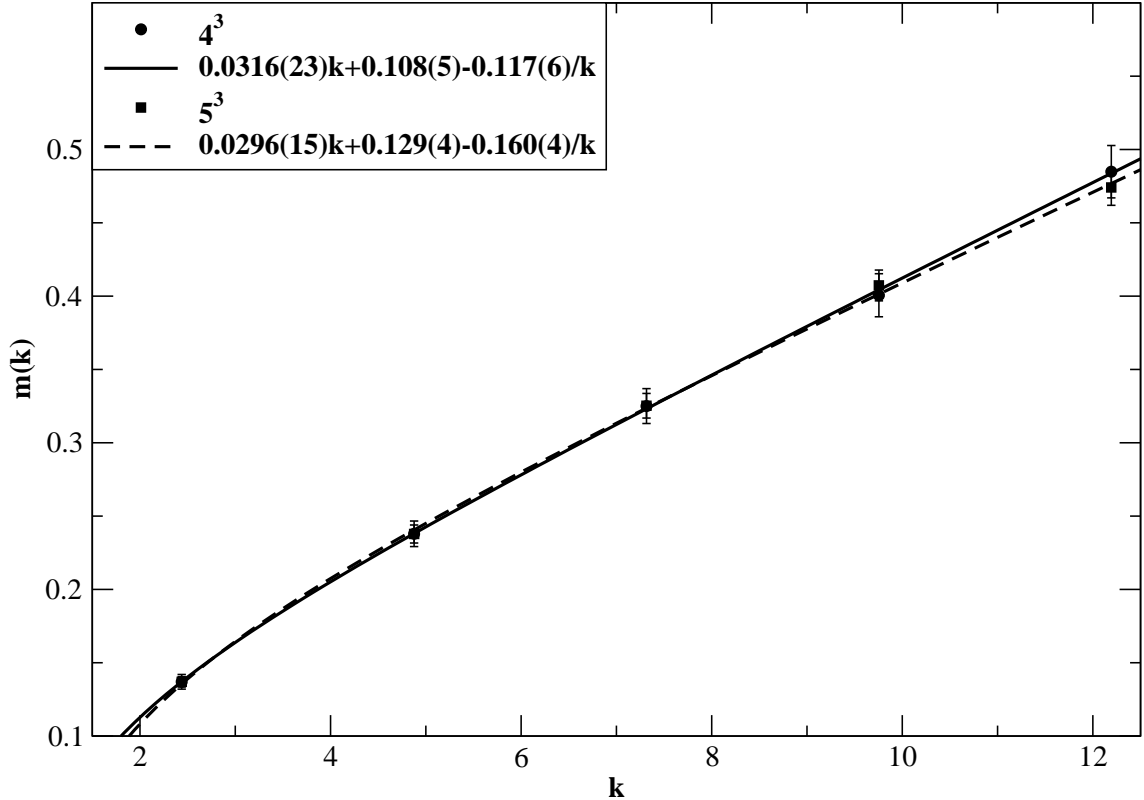
{ The ground state energy $m(k)$ as a 
function of $k$ on two different lattices 
at $b=0.8$.
\label{massl}
}}

\section{Creutz ratio}\label{crratio}

It is natural to ask how the Creutz ratio~\cite{Creutz:1984mg},
\be
\chi(K,J) = -\ln \frac{W(K,J)W(K-1,J-1)}{W(K,J-1)W(K-1,J)},
\ee
performs as an
observable from which to extract the string tension.
If we were to use Creutz ratios, we would have
smeared all links using all staples. But one can
still ask how the Creutz ratio behaves with the asymmetrically smeared 
links.
We show this for square loops ($J=K$) at $b=0.8$
and $\tau=2.5$ in Fig.~\ref{creutz}.
The solid lines show the estimate for the $\sqrt{\sigma}b_I$
as obtained from the analysis in this paper. There is no
evidence for a plateau in the Creutz ratio in the range of
$k$ shown in Fig.~\ref{creutz}. It is possible the situation
would be different if we had smeared all links. 

{\vskip 2cm
\EPSFIGURE
{creutz.eps, width=\textwidth}
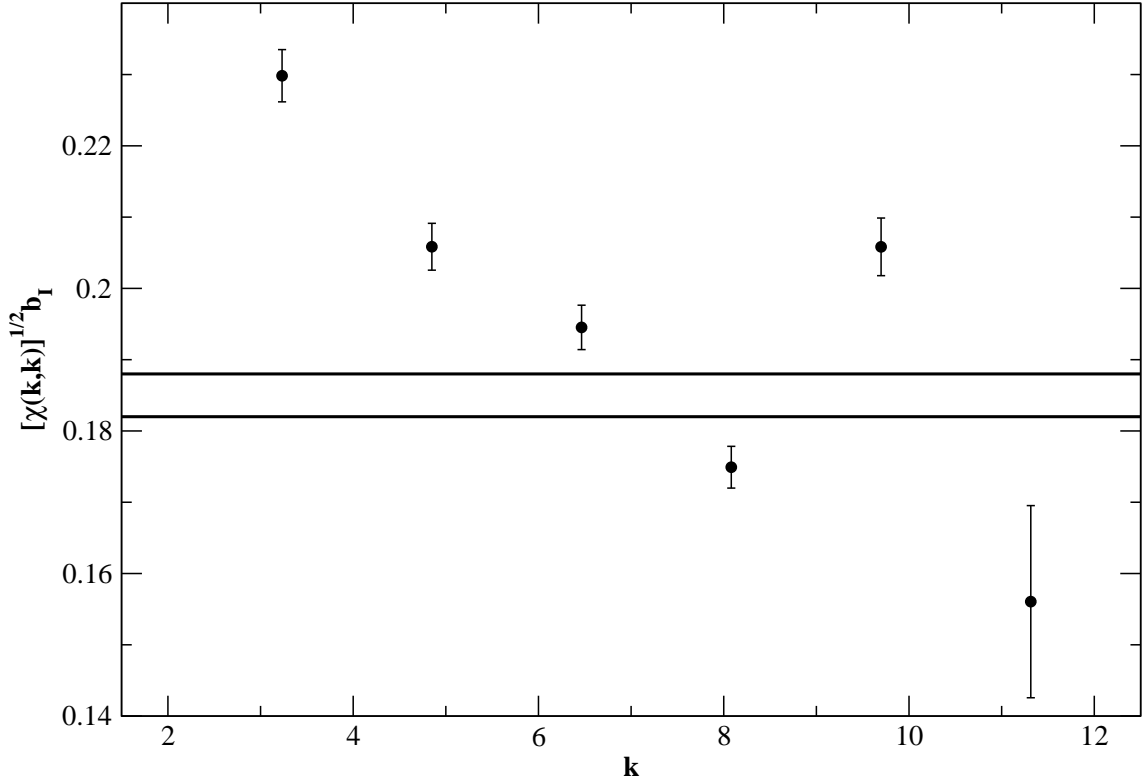
{Behavior of the Creutz ratio for square loops 
at $b=0.8$.
\label{creutz}
}}

Each data point in Fig.~\ref{creutz} is obtained using only
four different Wilson loops, {\em i.e.} four of the data points in 
Fig.~\ref{wloop1}.
This is quite different from
the analysis in this paper. Seven different Wilson loops
in Fig.~\ref{wloop1} are used to extract one $m(k)$ point in 
Fig.~\ref{tension}, and the loops used
for different $k$ form independent sets. Then the $m(k)$ are fit to 
determine the string tension. Both folded and unfolded loops contribute 
together. This is the main
reason we succeeded in extracting the string tension using
the range of Wilson loops considered here. To extract the string tension 
using Creutz ratios, larger loops
and therefore larger statistics and possibly larger $N$
would be needed.

\section{Conclusions}

We used Wilson loops with smeared space-like links and unsmeared
time-like links to obtain an estimate for the string tension in the
large $N$ limit of three dimensional Yang-Mills theory.  Invoking
large $N$ continuum reduction, we included Wilson loops larger than
the size of the lattice.  Since we used smeared space-like links, the
Wilson loops for fixed length in space and varying length in time
showed excellent agreement with a single exponential. The ground state
energy so obtained was fit using three parameters to get an estimate
for the string tension. The ground state energy exhibited short
distance behavior at the shortest length used in the paper but large
enough distances were used to get an estimate for the dimensionless
string tension with small errors (Equation~\ref{thisresult}).  These
results validate the method of continuum reduction for calculating
quantities based on the space-time dependence Wilson loops.

\acknowledgments

R.N. acknowledges partial support by the NSF under grant number
PHY-055375.

\end{document}